\documentclass[preprint,12pt]{elsarticle}

\usepackage{amssymb}
\usepackage{hyperref}
\usepackage{amsmath,amssymb}

\journal{ }

\begin{document}

\begin{frontmatter}



\title{DeepSafety:Multi-level Audio-Text Feature Extraction and Fusion Approach for Violence Detection in Conversations}



\cortext[a1]{Corresponding author}
\author{Amna Anwar}
\ead{amna.anwar@ntu.ac.uk}
\author{Eiman Kanjo}
\author{Dario Ortega Anderez}
\address{Smart Sensing Lab, Computer Science Department, Nottingham Trent University}

\begin{abstract}
Natural Language Processing has recently made understanding human interaction easier, leading to improved sentimental analysis and behaviour prediction. However, the choice of words and vocal cues in conversations presents an underexplored rich source of natural language data for personal safety and crime prevention. When accompanied by audio analysis, it makes it possible to understand the context of a conversation, including the level of tension or rift between people.  
Building on existing work, we introduce a new information fusion approach that extracts and fuses multi-level features including verbal, vocal, and text as heterogeneous sources of information to detect the extent of violent behaviours in conversations. Our multilevel multimodel fusion framework integrates four types of information from raw audio signals including embeddings generated from both \textit{BERT} and \textit{Bi-long short-term memory (LSTM)} models along with the output of \textit{2D CNN} applied to \textit{Mel-frequency Cepstrum (MFCC)} as well as the output of audio Time-Domain dense layer. The embeddings are then passed to three-layer \textit{FC} networks, which serve as a concatenated step. Our experimental setup revealed that the combination of the multi-level features from different modalities achieves better performance than using a single one with \textit{F1 Score=0.85}. We expect that the findings derived from our method provides new approaches for violence detection in conversations.

\end{abstract}




\end{frontmatter}


\section{Introduction} \label{sec:intro}
The multi-dimensional nature of speech and conversation presents a demanding task for computational machines to understand how individuals communicate with one another. The understanding of a conversation is highly subjective to the individuals involved, due to the merger of speech, communication patterns, and the surrounding context. Individuals take on different personalities by different triggers, and therefore there is a need to employ multiple modalities simultaneously. This challenging environment becomes even more complex when violent or harmful language is incorporated, as inaccurate detection could lead to potential consequences. 

Ultimately, violent acts consist of individual human beings inflicting harm on other human beings. The emotional states of the perpetrators at those moments can have a decisive effect on the degree of violence and even on whether any aggression occurs at all. On the other hand, aggression is a behaviour, emotion is a feeling state, and so the links between aggression and behaviour involve relationships between objective actions and subjective feelings \cite{Baumeister2003}. Several studies reveal that intimate-partner relationships can provide a potential context for intense emotions and conflicts, which can result in serious injuries or even death \cite{allen2007patterns}.

Emotion is naturally a complex and subjective research area, as a number of conversational, environmental, and personal factors can be combined to form the wider context of what is occurring. Given the ambiguity and versatility of personal emotions, their accurate recognition becomes an extremely challenging task which has been investigated by several researchers \cite{bhavan2019bagged,davidson2017automated}. Researchers have attempted to detect and recognise emotions using speech recognition \cite{bhavan2019bagged,li2019improved,atmaja2019speech}, image analysis \cite{du2019spatio,hu2019video,hajarolasvadi2020deep} and Natural Language Processing (NLP) which is often applied to the expansive datasets formed through the use of social media services such as Facebook, Twitter, and YouTube to understand the sentiment of the interaction \cite{batbaatar2019semantic,yang2020dacnn,plaza2020improved,joshi1991natural}. The ultimate objective of NLP is to interpret, decipher, and make sense of the human languages in a manner that is valuable. Therefore, emotion analysis including negative emotions and behaviour analysis might help in detecting the level of potential violence in human interaction. Recently, there have been multiple research efforts that looked at detecting negative emotions such as abusive language \cite{mossie2020vulnerable} and bullying \cite{sharma2018nlp} in online conversations, however, it is much more challenging to obtain a recorded conversation that is happening behind closed doors in real-life settings. Focussing entirely on text means the analysis misses the vocal cues and other sound clues available in the environment when the conversation occurs. Violence characterisation is subjective in nature, which makes violent incidents being usually manifested through characteristic audio signals (e.g.,  screams,  gunshots  etc). Sound (including low-frequency sounds) is often utilised in movies to elevate tension. Audio signals for violence detection are often used as an  additional  feature  where abrupt  changes in  energy level of the audio signal are detected using the energy entropy criterion \cite{Giannakopoulo06}. 

So far the field of multi-modal violence analysis in conversations has not received much attention. Inspired by these observations, we argue that with the power of technology and the ability to momentarily assess real-world data, it is now possible to go beyond sentimental analysis, hate, and abusive language detection online to detect violence during verbal conversations. To this end, we propose a novel approach to aspect-level violence classification based on multi-model data fusion. The main contributions of this work are three-fold:

\begin{enumerate}
    \item A multimodal fusion system for violence detection in conversations using features and embeddings automatically extracted with a combined deep learning approach.
    \item  A multi-level feature extraction approach that utilises transcribed text from audio conversation as an effective modality for detecting violence. To the authors’ knowledge, this is the first time that the combination of audio Frequency and Time Domain features combined by transcribed text about violence has been proposed.
    \item To motivate further research on violence detection using multimodel approaches in real-time using wearable and portable technologies that can assess the violence level and send an alert.
    
\end{enumerate}

Our data fusion approach developed at different levels as follows: 
\begin{enumerate}
  \item Multi-level data extraction, resulting in rich and heterogeneous fused audio and text data. 
  \item Feature Fusion by extracting key text features including (Psycho Linguistic Lexicon Linguistic Inquiry \cite{pennebaker2001linguistic} and Bidirectional Encoder Representations from Transformers (BERT) \cite{devlin2018bert}, and audio features using Mel-frequency Cepstrum (MFCC) and Time Domain features.
  \item Decision fusion by combining multiple classifiers from different modalities for violence detection.
\end{enumerate}

Thus, to provide objective metrics of feature relevance and system performance in full realism, a number of relevant recordings from english television series and other audiovisual related material containing violence or abusive conversations are retrieved. The corresponding audio signals are extracted from the videos and segmented into 10ms chunks and labelled according to whether they contain verbal abuse or not on a scale from 1 to 5.  The segments are then fed into the fusion algorithms (both audio and NLP features after transcribing the clips). The resultant dataset embodies a total of 3 hours and 36 minutes of audio data. With this, a total of 1295 segments are obtained, from which 633 contain verbal abuse and the remaining 662 do not contain any form of verbal abuse. The experimental results show that our approach has performed well with our fusion model achieving nearly F1 score=0.85, which contributes towards the advancement of violence detection techniques in conversations.

The remainder of the paper is organised as follows. We present related work on the issue in the following section. Then we explain and analyse the study used in this paper. Next, the attained results along with their implications and limitations are elaborated. Finally, we conclude the paper and suggest some potential future work.

\section{Vocal Aggression Detection}
Over the past few years there have been many research efforts to recognise emotions and sentimental analysis in text \cite{medhat2014sentiment}. Very little attempts has been found that are linked to detecting violence in language. These attempts were limited to hate speech \cite{watanabe2018hate} or violence on social networks \cite{cano2013weakly}.
The majority of the sentiment analysis research has focused on textual data \cite{zhang2018deep}. However, two main factors are leading to the employment and development of alternative approaches and methods for emotion inference. On the one hand, the advent of ubiquitous computing is leading to an increasing use of ubiquitous sensors such as microphones or video cameras which can be easily and conveniently embedded within different environments. On the other hand, the growth in popularity of social media such as YouTube, Instagram or Facebook has led to an increasing number of users expressing their feelings and views in audio, visual, and audiovisual formats \cite{poria2017review}.

Traditionally, the relevant authorities have relied on video-only solutions, but crime prevention technologies continue to shift and adapt to a new standard of sight and sound for an effective solution. Deploying proximity detection, audio, and other sensors analytic  gives the relevant authorities the means to better recognize potential risk or threat. Specifically, vocal aggression detection technology is gaining traction in security system installations, as it offers staff better situational awareness, additional information about an event in progress, and faster opportunities for intervention. Technologies such as audio analytics not only offer an additional layer of evidence for responding to incidents \cite{gemmeke2017audio}, but also allow security personnel to assess the tone of a situation in real time and take immediate action before circumstances deteriorate. Similarly to how the human ear processes audio, sound detection software analyses acoustic events through advanced algorithms and classifies them into predetermined categories.

Aggression detectors are capable of accurately recognizing stress and duress in a person’s voice, automatically and objectively detecting the presence of aggression, anger, or fear, and warn individuals by a visual alert or an audible alarm. For example, consider a conversation between a victim and known offender and an argument ensues, an audio analytic device equipped with aggression detecting software could quickly identify the raised level of stress in either individual’s voice and immediately flag the area for immediate help. The major benefit of the aggression detection, proximity detection, and audio analytics solutions on the Edge device comes from the fact that the technology maintains individual’s right to privacy. Sound detectors do not listen for speech, language, or key words, but instead analyse frequencies, volume level, time duration, and other acoustic patterns \cite{giannakopoulos2006violence}.

\subsection{Text Representation}\label{subsec:rev:Text}
The majority of research surrounding sentiment analysis is focused on detecting negative and positive sentiments on data collected from social media platforms such as Facebook, Instagram and Twitter, being one of the prominent sources \cite{ortigosa2014sentiment}\cite{naf2019sentiment} \cite{severyn2015twitter}. In recent years, sentiment analysis has gained immense recognition in the literature, as the analysis of an individual's emotional state and its dynamics can provide research with queues that could be used for predicting personality and speech patterns. These predictions can be used as a form of violence detection in different scenarios.

Sentiment analysis has become more mature in the recent decade \cite{HUSSEIN2018330}  and the most commonly deployed classification techniques were SVM, Naive Bayes and Maximum Entropy which are based on the word bag model. The word bag model however disregards the sequence of the words in a sentence, which can have a significant effect on the meaning of the sentence as well as change the sentiment as discussed in the survey carried out by \cite{CHATURVEDI201865}.

The detection of violent and offensive language detection is conventionally classed in specific types; such as the detection of bullying as seen in \cite{zhao2016automatic}, identification of aggression \cite{chen2011detecting} and hate speech identification \cite{chen2019complementary} . NLP has been applied to a great extent in studies surrounded sentiment analysis to exploit the lexical syntactic features of phrases and sentences to detect offensive language \cite{chen2012detecting} . 
As the NLP community grows and becomes increasingly popular for the automatic detection of hate speech, offensive abusive language, there are common patterns that can be seen. Initially, data goes through pre-processing to gain useful insights and to clean up the text from any irrelevant information using methods such as removing Punctuation's, Stopwords,Tokenization, parts of Speech Tagging, and Lemmatization \cite{cambria2017sentiment}. Machine learning based classifiers were researched in the early days \cite{warner2012detecting}\cite{burnap2015cyber} to detect abusive language; feature-based linear classifiers (REF 1 2), neural network architectures such as convolutional neural network (CNN) or LSTM which is later then fine tuned by pre-trained language models such as BERT and Elmo.  

However, existing word embedding methods \cite{peng2019transfer}, which use a limited window size, cannot exploit semantic information in the global context. Additionally, such an algorithm transforms a word into a stable vector. As a result, the vector is unable to accurately represent its context at different locations. 

Recently, context-specific language representation models, such as ULMFiT \cite{howard2018universal}, OpenAI GPT \cite{qu2020text}, ELMo \cite{peng2019transfer} and BERT \cite{alatawi2021detecting}, have been designed by jointly conditioning the left and right context with deep neural networks. Furthermore, these models can dynamically adjust the word vector according to its context. Our work mainly relies on BERT, which provides us with a strong baseline, and we further modify its standard network structure to incorporate the target information.

\subsection{Inference from Audio}\label{subsec:rev:Audio}
The research field of emotion recognition using audio-based technologies is gaining increasing attention in recent years, leading to an increasing number of research studies in the field \cite{kaushik2013sentiment,maghilnan2017sentiment,luo2019audio,li2019acoustic,rane2020audio}. For instance, the work in \cite{atmaja2019speech} proposes a Bidirectional Long Short-Term Memory (Bi-LSTM) with attention model to classify four different emotions including "anger", "Excitement", "Neutral" and "Sadness" from the IEMOCAP dataset, which includes a range of dyadic sessions where actors perform  improvisations or scripted scenarios specifically selected to elicit emotional expressions. To do so, a silence removal signal processing step followed by the extraction of a 34-dimensional feature vector composed of a range of time domain, frequency domain, MFCCs and Chroma-based features is implemented. The results reported outline a maximum classification accuracy of 70.34\%. Making use of the same dataset,  \cite{li2019improved} proposes a self-attentional CNN-BLSTM classification model fed with the Mel-spectrograms of the corresponding audio files to classify between the four above-mentioned emotions. A 81.6\% classification accuracy is achieved by the above methodology. In \cite{yenigalla2018speech} a multi-channel CNN fed with Mel-spectrograms and phoneme embeddings generated by the word2vec model introduced in \cite{mikolov2013efficient}. 

A very limited body of literature has concerned the recognition of violent and aggressive speech from conversations (audio). Therefore, this work aims at filling this gap by providing a solution for aggressive behaviour that uses sound as one of the inputs. 

\subsection{Inference through Data Fusion}\label{subsec:rev:Fusion}
The multimodal fusion approach has been used for emotion recognition using audio, video, and physiological data \cite{povolny2016multimodal}. Researchers have incorporated audio, visual and text information as 90\% of the existing literature makes use of them for a more accurate multimodal affect \cite{poria2017review} \cite{perez2013utterance} \cite{wollmer2013youtube} \cite{poria2015deep} \cite{yoon2018multimodal}, \cite{buitelaar2018mixedemotions}. \cite{poria2017review} main focus was to advocate the multimodal over a unimodal approach. 

Gaurav Sahu proposed a feature engineering-based approach to tackle speech emotion recognition, similar to the attempted approach they first trained all of their audio and text features separately and they then fused them at the feature vectors which were concatenated \cite{sahu2019multimodal}.
Feiyang C et al. approached the multimodal sentiment analysis by incorporating audio and text. Their strategy was to use both multi-feature fusion and multi-modality fusion to improve the accuracy of audio-text sentiment analysis. Their focus was solely on general sentiment rather than the detection of hateful or negative sentiment and therefore they used the CMU-MOSI dataset and called their Deep Feature Fusion - Audio and Text Modality Fusion (DFF-ATMF) model \cite{chen2012detecting}. 
Moises H. R. Pereira et al. on the other hand, utilises audio, textual, and visual clues that they extracted from news videos. They applied state-of-the-art computational methods to automatically recognise emotion from facial expressions; the extraction of modulations from the speeches of the participants and the sentiment analysis on the closed captions of the videos. They used a dataset that contained 520 annotated news videos from three famous TV newscasts and reported an accuracy of up to 84\% for the sentiment classification task \cite{pereira2016fusing}. 

\section{Dataset}

\subsection{Experimental Setup and Dataset}\label{subsec:exp_setup}

For the implementation of the system, a number of relevant videos from television series and other audiovisual related material containing domestic abuse scenes with violent and abusive conversations are retrieved and the corresponding audio signals are extracted from the videos. A posteriori, a total of 1295 audio files are divided into time segments with a length of 10 seconds and labelled according to whether they contain verbal abuse (including violence) or not on a scale from 1 to 5. They are derived from incongruous British television series and the different elements of the proposed dataset are described as follows:

\begin{enumerate}
    \item A collection of 50 minutes and 42 seconds of video data from scenes from the series Eastenders \cite{youtube}. With this, 304 segments of audio with a duration of 10 seconds are obtained, 195 not containing verbal abuse, and the remaining 110 segments containing verbal abuse. Another 133 10 second segments were retrieved from 4 to 7 minute long YouTube videos. Within these, 73 were non violent and leaving 60 with violence. 
    
    \item A set of 507 audio files derived from 5 to 10 minute videos from the British TV serial Coronation Street \cite{youtube} were collected and split into 10 second segments of audio files. From these, 352 do not contain any verbal abuse leaving 288 segments with verbal abuse.
    
    \item The corpus also includes scenes that showcases violence and abuse from another popular series named Emmerdale. From this series there are a total of 350 10 second audio clips from which 115 did not contain any violent language, leaving 235 with abusive language. 
    
\end{enumerate} 

The resultant dataset embodies therefore a total of 1295 segments, from which 633 contain verbal abuse and the remaining 662 do not contain any form of verbal abuse. Implementation was primarily carried out on Google Colab as it provides a single 12GB NVIDIA Tesla K80 GPU that can be used for up to 12 hours continuously. 
The Python library “ibmwatson”was used to process the video files into 10 seconds segmented audio files. 

The reason for limiting the size of each segment to 10 sec to provide a consistent functionalities across the audio and text features. while longer segments contain more information, violent and non0violent events will be mixed. To convert audio to text two approaches were used to eliminate any errors. The first approach was “SpeechToTextV1” which transcribed the audio files into text; however, due to the British accent the transcription was incorrect and had to be looked over by the labellers as they labelled the text.

We choose this dataset as it was not possible to obtain recordings of real-life conversations given the nature and the sensitivity of the contents.

\subsection{Data Pre-Processing}\label{subsec:-audiopreprocess}

Speech recognition, or speech-to-text, is the computational process of identifying spoken words within an audio signal and preserving those words in readable text. To enable the accurate recognition of words and the subsequent translation to text, the audio files need to go through various key pre-processing steps. These include the removal of unwanted background noise that does not correspond to human speech, as well as the segmentation of the audio signal and the grouping of homogeneous regions of the audio signal regarding the speaker identity (speaker diarization). With this, we aim to obtain $N$ "noise-free" audio signals corresponding to the speech of the $N$ speakers taking part in a conversation.

\subsection{Data Labelling and Annotation}\label{subsec:-audiopreprocess}

\begin{table}
\begin{tabular}{|p{0.5cm}|p{8cm}|p{0.5cm}|p{0.5cm}|p{0.5cm}|p{1cm}|}
 \hline
 ID & Transcript & R1 & R2 & R3 & Label \\
 \hline
 030 & "are you suggesting i don't go? well this
is gonna make you feel uncomfortable
uh i don't know i've got people coming
in for a few days at home will do good
" &1 &2 &1 & 2 \\
\hline
 001 & "whatever hey hey hey hey no tip allows you to do
that my friend take your enough for data
take your hands off my husband
"  & 3&2 & 3& 3\\
 \hline
 061 & "something else wouldnt it? wouldnt it? Cause there's always something else isn't it. I cant live like this. No....no  Chantelle please.... look i just
"  & 2&3 &3 &3 \\
 \hline
\end{tabular}
\caption{An extract of the data set with the three independent reviews and the determined label.} \label{table:dataset}
\label{table:1}
\end{table}
In machine learning, data labelling is the process of adding one or more meaningful and informative labels to pre-identified data to provide context so that a machine learning model can learn from it.
The common approach in language processing to engineer labels is keyword-based hate speech detection using ready-made lexicons \cite{chen2012detecting} \cite{sood2012using}. Since there are no common or agreed lexicons for violent words, the data was collected and labelled by three reviewers to suit the needs of the study. Although by utilising lexicons such as the HateBase,  the results of the systems do perform high, it was decided not to be used as they can be highly biased and unreliable. Furthermore, it is challenging to maintain and keep the lists up to date \cite{macavaney2019hate}. 

Waseem et al.\cite{waseem2016hateful} study was applied as a guideline as they made a generic definition based on the hate related content found on social media to address the problem of detecting it on Twitter. The Gender Studies and Critical Race Theory (CRT) are used as baseline. For their study they attempted to annotate a total of 16,849 tweets corpus into three categories: “Racism, Sexism and Neither” by themselves. To ensure the corpus was reliable and impartial, they had a “a 25 years old woman studying gender studies and non-activist feminist to reduce annotator bias” \cite{mozafari2020hate}.

For this work, three labellers were recruited to annotate each transcribed speech segment; one of them being a graduate in Sociology who had a focus on domestic abuse. The labellers gave a score to each segment within the dataset with a violence intensity between 0-5 on a Likert scale.

\begin{figure}
    \centering
    \includegraphics[width=0.7\textwidth]{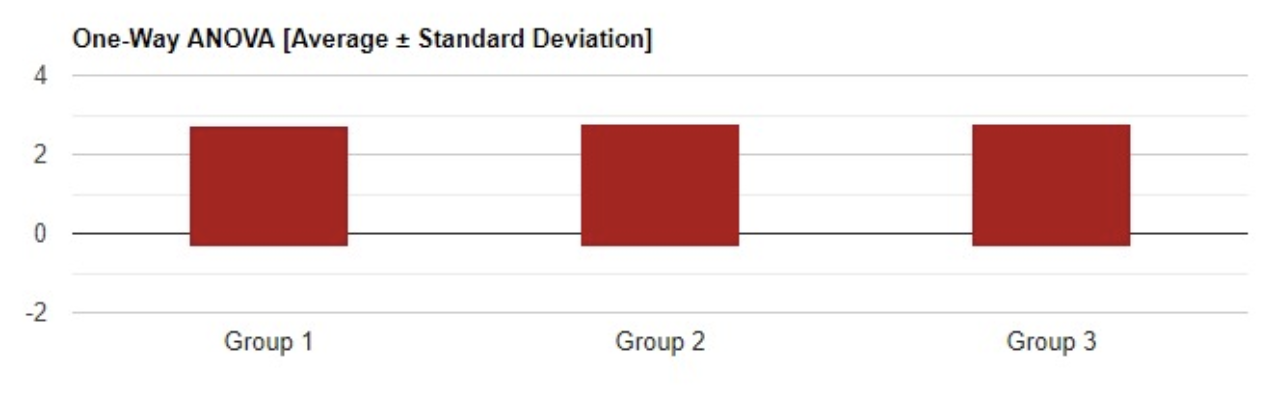}
    \caption{Average and Standard Deviation values of the three different reviewers ranking. }
    \label{fig:anova}
\end{figure}

A posteriori, the resultant segments' labels were obtained by averaging the scores given by the three labellers as the sentiment polarity. The three labellers were used to mitigate the natural subjectivity and complex feelings of language, and define mean values to be used as a curated set of resultant scores. To test the reliability of the scores, a one-way analysis of variance (ANOVA) test was completed and found no significant difference in the mean rankings provided (p = 0.932 $>$ 0.001). The mean ranking for each reviewer was 1.21, 1.23 and 1.22 with similar standard deviations as presented in figure \ref{fig:anova}.

\section{Features Extraction} \label{sec:methods}
We leverage two types of features to develop the violent language recognition system, including audio and text features. Features extraction methods for both types will be described in this section.

Using Google’s language model Bidirectional Encoder Representations from Transformer (BERT) \cite{devlin2018bert}, we generated word embeddings of each transcribed audio segment. BERT incorporates a method called a masked language model, which implements a special classification token (CLS) at the start and a special separation token (SEP) at the end. It also enables the randomisation of some words which are then substituted by a (MASK) in the sequence.

\subsection{Text Pre-processing and Feature Extraction}\label{subsec:textpreprocess}
\begin{figure}
    \centering
    \includegraphics[width=0.5\textwidth]{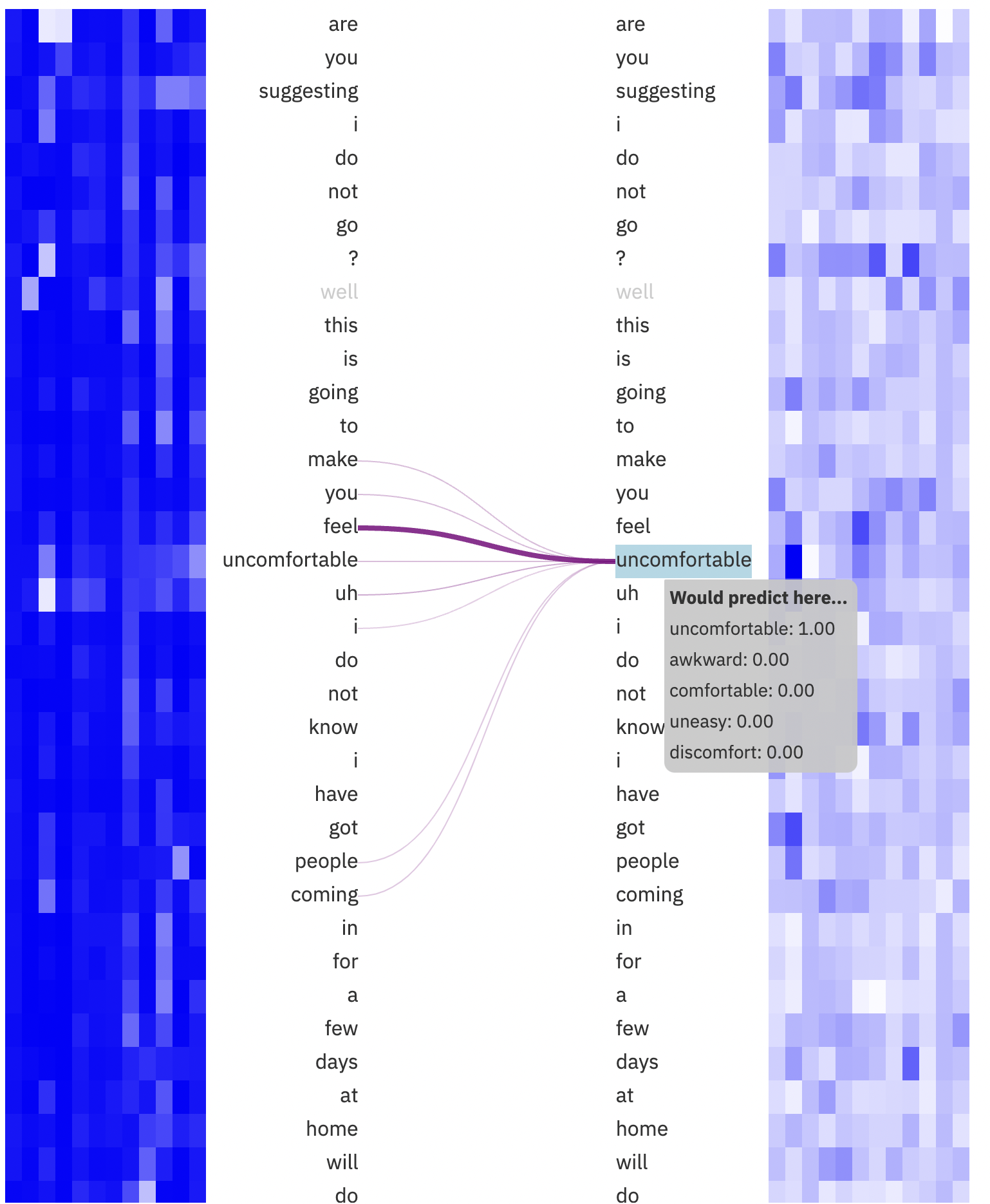}
    \caption{Output visualization of attention between example selected words from the dataset. }
    \label{fig:wordpredictionheatmap}
\end{figure}

BERT utilizes transformer, a mechanism of awareness which understands contextual aspects of speech in a single text. In its simple form, transformer contains two different mechanisms – the encoder which reads the text input and the decoder which generates a prediction for the task. When implementing the BERT framework, the base model consists of a 12 layer transformer block encoder. With each block of the encoder it comprises a 12 head self-attention layer and a 768-dimensional hidden layer, consisting a total of 110M parameters. The general model enables inputs up to a series of 512 tokens and outputs word embeddings. 
One of the key steps in data mining and knowledge discovery is feature extraction as the goal is to select the most suited features $f_j$ to enable more accurate accuracy at the classification stage. Figure \ref{fig:wordpredictionheatmap}, shows an example of  of attention between selected words in the input. The darkness of the colour indicates the strength of the attention weight (some attention weights have very low values so they are invisible).

\begin{figure}
    \centering
    \includegraphics[width=1.2\textwidth]{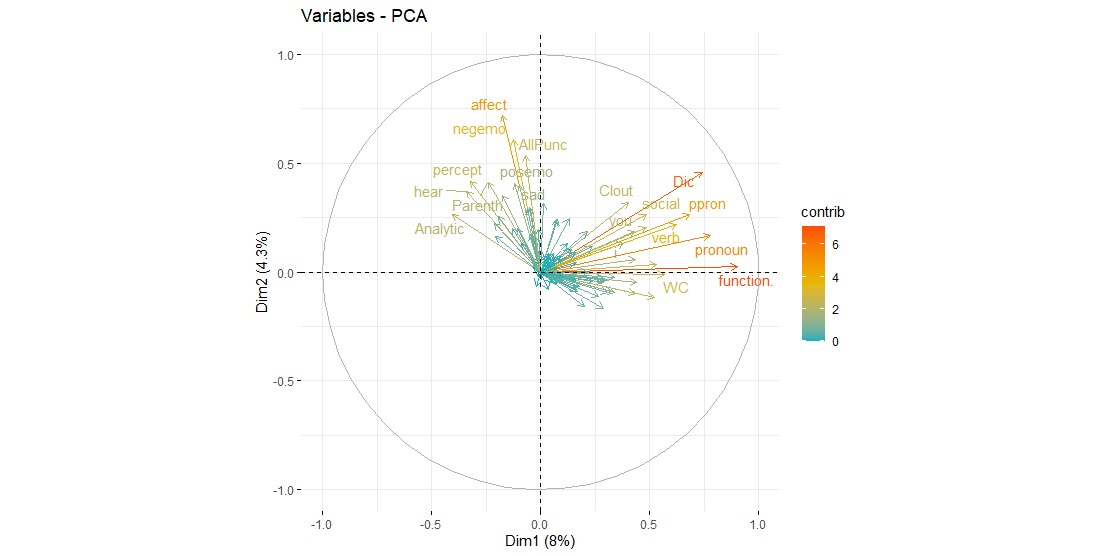}
    \caption{PCA analysis on LIWC features}
    \label{fig:pca-new}
\end{figure}

\begin{figure}
    \centering
    \includegraphics[width=1.2\textwidth]{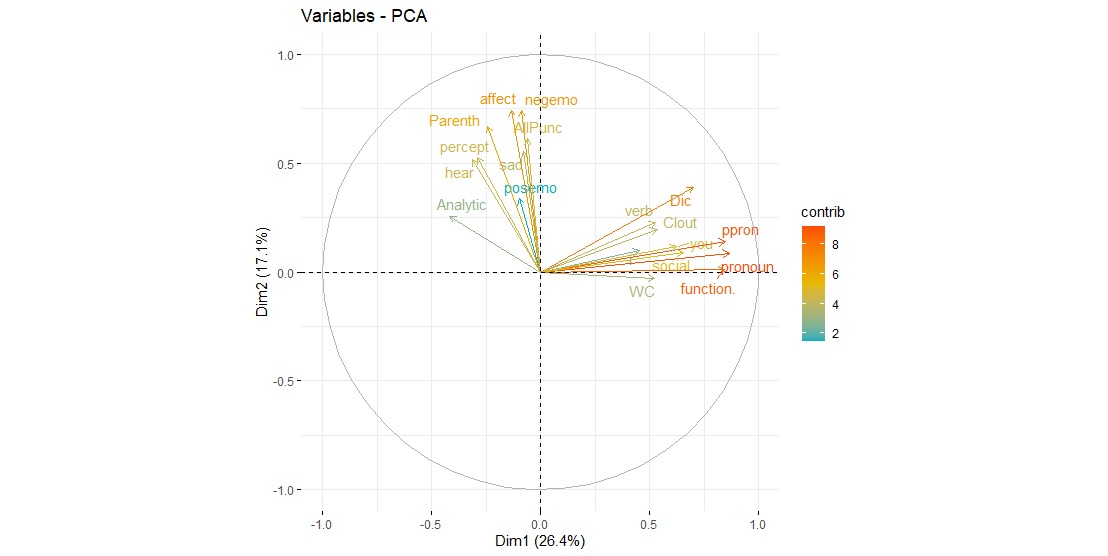}
    \caption{PCA revised analysis on LIWC features}
    \label{fig:pca-new}
\end{figure}

To extract features from the dataset, our model makes use of the psycho-linguistic lexicon Linguistic Inquiry and Word Count (LIWC) package \cite{Pennebaker15}. The purpose of the package is to differentiate the semantic-syntatic patterns and the different contextual information between of abuse and advice. It is a lexicon that was created by psychologists with a focus on determining different emotional cognitive, and linguistic components from text. We implemented LIWC to enable the extraction of the contextual parameters for each transcription segment..
To supplement the extraction of linguistic features, we utilise Principal Component Analysis (PCA) \cite{Jolliffe90} to distinguish the highest contributing factors across the dataset. PCA is a type of dimensionality reduction technique, which enhances the interoperability of large multidimensional data. Figure \ref{fig:pca-new}, presents the highest contributing variables which includes parameters such as functional words (to, very, it), analytical thinking and common verbs (carry, think, go). To check the level of correlation between there variables a Bartlett's test of sphericity was conducted. The p-value calculated was $<$0.1 which confirmed that PCA was appropriate. Through using PCA, it was possible to reduce the dimensionality of the dataset, while preserving the variance and significance of the  contributing variables.

\subsection{Audio Pre-processing and Features Extraction}\label{subsec:-audiopreprocess}
As mentione above, the original sampling rate of the different audio signals that compose the proposed dataset is 22050Hz. However, lower sampling rates can lead to higher classification rates since the far end of the spectrum is not expected to contain relevant information for speech-related applications. Given this, the performance of the system was studied across different sampling rates (22,050Hz, 16,000Hz and 11,000Hz) through the corresponding down-sampling of the audio signals.    

Unlike with other sensory signals or data formats such as inertial signals from Inertial Measurement Units (IMUs), physiological signals or images, the extraction of patterns from raw audio data typically involves the extraction of hand-crafted features to achieve competent performances. This is mainly due to the fact that raw audio data naturally comes in the form of time series (time domain), while substantial information about the audio signals can be only be revealed within the frequency domain. In line with this, this paper proposes a feature vector that incorporates a wide array of self-engineered features, revealing relevant information about the signal in both the time and the frequency domains. Such features are defined in Section \ref{sub:sub:TimeDomain} (time domain), Section \ref{sub:sub:FrequencyDomain} (frequency domain) and Section \ref{sub:sub:TimeFreqDomain} (time-frequency domain).   
\subsubsection{Time Domain}\label{sub:sub:TimeDomain}
The array of features calculated in the time domain can be defined as follows. 

\begin{itemize}

\item [-] Amplitude envelope:

\begin{equation}
    AE_t=\max_{k=tK}^{(t+1)K-1} s(k)
\end{equation}

\noindent where $t$ refers to the $t^{th}$ frame, $K$ is the frame size and $s(k)$ is the amplitude of the $k^{th}$ sample of the signal.   

\item [-] Amplitude envelope discrete derivative:
\begin{equation}
    \Delta AE_t = AE_t - AE_{t-1}
\end{equation}

\noindent where $AE_t$ is the amplitude envelope of frame $t$ and $AE_{t-1}$ is the amplitude envelope of frame $t-1$.

\item [-] Root mean square energy:
\begin{equation}
    RMS_t=\sqrt{\frac{1}{K}\sum_{k=tK}^{(t+1)K-1}s(k)^2}
\end{equation}

\noindent where $t$ refers to the $t^{th}$ frame, $K$ is the frame size and $s(k)$ is the amplitude of the $k^{th}$ sample of the signal.

\item [-] Root mean square discrete derivative:
\begin{equation}
    \Delta RMS_t = RMS_t - RMS_{t-1}
\end{equation}
\noindent where $RMS_t$ is the root mean square energy of frame $t$ and $RMS_{t-1}$ is the root mean square energy of frame $t-1$.

\item [-] Zero crossing rate:
\begin{equation}
    ZCR_t = \frac{1}{2}\sum_{k=tK}^{(t+1)K-1} \left|sgn(s(k))-sgn(s(k+1))\right| 
\end{equation}
\noindent where $t$ refers to the $t^{th}$ frame, $K$ is the frame size and $sgn(s(k))$ is the sign of the amplitude of the signal at sample $k$. 

\item [-] Zero crossing rate discrete derivative:
\begin{equation}
    \Delta ZCR_t = ZCR_t - ZCR_{t-1}
\end{equation}
\noindent where $ZCR_t$ is the zero crossing rate of frame $t$ and $ZCR{t-1}$ is the zero crossing rate of frame $t-1$.

\end{itemize}

Following the computation of the above time domain features, basic descriptive statistics, including the mean, the maximum, the minimum, the standard deviation and the root mean squared from each of the features are calculated, leading to a 30-dimensional feature vector (6 features x 5 descriptive statistics) presented in Table \ref{tab:time_domain}.

\begin{table}[]
\caption{Features in the time domain.}
\centering
\resizebox{\columnwidth}{!}{%
\renewcommand{\arraystretch}{1.5}
\begin{tabular}{|l|l|l|l|l|l|}
\hline
\textbf{Feature}   & \textbf{Mean}    & \textbf{Max}    & \textbf{Min}    & \textbf{STD}    & \textbf{RMS}    \\ \hline
\textbf{AE}        & mean\_ae         & max\_ae         & min\_ae         & std\_ae         & rms\_ae         \\ \hline
\textbf{delta AE}  & mean\_delta\_ae  & max\_delta\_ae  & min\_delta\_ae  & std\_delta\_ae  & rms\_delta\_ae  \\ \hline
\textbf{RMS}       & mean\_rms        & max\_rms        & min\_rms        & std\_rms        & rms\_rms        \\ \hline
\textbf{delta RMS} & mean\_delta\_rms & max\_delta\_rms & min\_delta\_rms & std\_delta\_rms & rms\_delta\_rms \\ \hline
\textbf{ZCR}       & mean\_zcr        & max\_zcr        & min\_zcr        & std\_zcr        & rms\_zcr        \\ \hline
\textbf{delta ZCR} & mean\_delta\_zcr & max\_delta\_zcr & min\_delta\_zcr & std\_delta\_zcr & rms\_delta\_zcr \\ \hline
\end{tabular}
\label{tab:time_domain}
}
\end{table}

\subsubsection{Frequency Domain}\label{sub:sub:FrequencyDomain}
Alongside the features extracted in the time domain, a number of features are extracted in the frequency domain. To do so, the time series $x$, corresponding to the audio signal, is converted into the frequency domain using the Short Time Fourier Transform (STFT) along with a Hann window as follows:
\begin{equation}
    S(m,k)=\sum_{n=0}^{N-1}x(n+mH)w(n)e^{-i2\pi n\frac{k}{N}}
\end{equation}
\noindent where $m$ refers to the $m^{th}$ frame or temporal bin, $k$ refers to the $k^{th}$ frequency within the frequency bins, $N$ is the frame size, $H$ is the hop size and $w(n)$ is the Hann Window function applied to the $n^{th}$ sample within a frame $m$. The Hann Window is given by:
\begin{equation}
    w(k) = 0.5(1-cos(\frac{2\pi k}{K-1})), k = 1...K
\end{equation}

Once the signal is converted into the frequency domain and the windowing function is applied to each signal frame, the following features are extracted:

\begin{itemize}

\item [-] Band energy ratio (BER):
\begin{equation}
    \Delta BER_t = \frac{\sum_{n=1}^{F-1} m_t(n)^2}{\sum_{n=F}^N m_t(n)^2}
\end{equation}
\noindent where $m_t(n)^2$ is the power of the signal at frame $t$ and frequency bin $n$, $F$ is the split frequency and $N$ is the highest frequency bin within a frame $t$.   

\item [-] Band energy ratio discrete derivative:
\begin{equation}
    \Delta BER_t = BER_t - BER_{t-1}
\end{equation}
\noindent where $BER_t$ is the band energy ratio of frame $t$ and $BER{t-1}$ is the band energy ratio of frame $t-1$.

\item [-] Spectral centroid (SC):
\begin{equation}
    SC_t=\frac{\sum_{n=1}^N m_t(n)n}{\sum_{n=1}^N m_t(n)}
\end{equation}
\noindent where $m_t(n)$ is the magnitude of the signal at the $n^{th}$ frequency bin and $N$ is the highest frequency bin within a frame $t$.

\item [-] Spectral centroid discrete derivative:
\begin{equation}
    \Delta SC_t = SC_t - SC_{t-1}
\end{equation}
\noindent where $SC_t$ is the spectral centroid of frame $t$ and $SC{t-1}$ is the spectral centroid of frame $t-1$.

\item [-] Spectral bandwidth (SBW):
\begin{equation}
    SBW_t=\frac{\sum_{n=1}^N \left|n-SC_t\right|m_t(n)}{\sum_{n=1}^N m_t(n)}
\end{equation}
\noindent where $m_t(n)$ is the magnitude of the signal at the $n^{th}$ frequency bin in frame $t$, $N$ is the highest frequency bin within a frame $t$ and $SC_t$ is the spectral centroid at frame $t$.

\item [-] Spectral bandwidth discrete derivative:
\begin{equation}
    \Delta SBW_t = SBW_t - SBW_{t-1}
\end{equation}
\noindent where $SBW_t$ is the spectral bandwidth of frame $t$ and $SBW{t-1}$ is the spectral bandwidth of frame $t-1$.

\item [-] Spectral roll-off:
\begin{equation}
    \sum_{n=0}^{R_{n-1}}\left|m_t(n)\right|=0.85\sum_{n=0}^{N-1}\left|m_t(n)\right|
\end{equation}
where $m_t(n)$ is the magnitude of the signal at the $n^{th}$ frequency bin in frame $t$, $N$ is the highest frequency bin within a frame $t$ and $R_{n-1}$ is the roll-off frequency bin.

\item [-] Spectral roll-off discrete derivative:
\begin{equation}
    \sum_{n=0}^{R_{n-1}}\left|m_t(n)\right| - \sum_{n=0}^{R_{n-1}}\left|m_{t-1}(n)\right|
\end{equation}

\item [-] Spectral Flux (SF)
\begin{equation}
    SF_t = \sum_{n=0}^{N-1} s(k,i) – s(k-1,i)
\end{equation}

\item [-] Spectral flux discrete derivative:
\begin{equation}
    \Delta SF_t = SF_t - SF_{t-1}
\end{equation}
\noindent where $SF_t$ is the spectral flux of frame $t$ and $SF{t-1}$ is the spectral flux of frame $t-1$.

\end{itemize}

\begin{table}[]
\caption{Features in the frequency domain.}
\centering
\resizebox{\columnwidth}{!}{%
\renewcommand{\arraystretch}{1.5}
\begin{tabular}{|l|l|l|l|l|l|}
\hline
\textbf{Feature}   & \textbf{Mean}    & \textbf{Max}    & \textbf{Min}    & \textbf{STD}    & \textbf{RMS}    \\ \hline
\textbf{BER}       & mean\_ber        & max\_ber        & min\_ber        & std\_ber        & rms\_ber        \\ \hline
\textbf{delta BER} & mean\_delta\_ber & max\_delta\_ber & min\_delta\_ber & std\_delta\_ber & rms\_delta\_ber \\ \hline
\textbf{SC}        & mean\_sc         & max\_sc         & min\_sc         & std\_sc         & rms\_sc         \\ \hline
\textbf{delta SC}  & mean\_delta\_sc  & max\_delta\_sc  & min\_delta\_sc  & std\_delta\_sc  & rms\_delta\_sc  \\ \hline
\textbf{SBW}       & mean\_sbw        & max\_sbw        & min\_sbw        & std\_sbw        & rms\_sbw        \\ \hline
\textbf{delta SBW} & mean\_delta\_sbw & max\_delta\_sbw & min\_delta\_sbw & std\_delta\_sbw & rms\_delta\_sbw \\ \hline
\textbf{SRO}       & mean\_sro        & max\_sro        & min\_sro        & std\_sro        & rms\_sro        \\ \hline
\textbf{delta SRO} & mean\_delta\_sro & max\_delta\_sro & min\_delta\_sro & std\_delta\_sro & rms\_delta\_sro \\ \hline
\textbf{SF}        & mean\_sf         & max\_sf         & min\_sf         & std\_sf         & rms\_sf         \\ \hline
\textbf{delta SF}  & mean\_delta\_sf  & max\_delta\_sf  & min\_delta\_sf  & std\_delta\_sf  & rms\_delta\_sf  \\ \hline
\end{tabular}
\label{tab:freq_domain}
}
\end{table}

Following the computation of the above frequency domain features, as with the features extracted in the time domain, basic descriptive statistics, including the mean, the maximum, the minimum, the standard deviation and the root mean squared are calculated from each of the features, leading to a 50-dimensional feature vector (10 features x 5 descriptive statistics), as presented in Table \ref{tab:freq_domain}.

\subsubsection{Mel-based Features }\label{sub:sub:TimeFreqDomain}

Mel-frequency cepstral coefficients are commonly used in automatic speech recognition \cite{784104}. Here, we briefly describe the MFCC method. First, the audio signal is segmented into short frames with some overlap. The reason for keeping the frames short is that the audio signal is assumed to be stationary over short time. The power spectrum of each frame is calculated using the periodogram.  Then  a mel-filter bank of triangular filters is applied to the power spectra, and energy in each filter is summed up. To match the features closely to human hearing, the logarithms of all the filter bank energies are computed. As the filter banks are usually overlapping in the frequency domain, a discrete cosine transform (DCT) is applied to the log filter bank energies. In the end, a set of lower DCT coefficients is taken which  represents the  MFCCs.  In order to exploit the complete discriminative ability of MFCC, first- and second-order derivatives of MFCC features are also used.

\section{Model}\label{subsec:methods:audio}
Given that the extracted features exhibit differing formats, different models are proposed regarding the resultant structure of the different feature sets.    
Our approach consists of 2D CNN Model and a Bi-LSTM with attention layer. First we feed the MFCC features and the audio Time Domain features though 3 layers of 2D Convolutions Neural Network (CNN) followed by Fully Connected (FC) layer. Then, we setup Bi-LSTM model for the LIWC features after applying PCA for dimensionality reduction.  
We utilise BiLSTM similar to \cite{9231008}, as it's often applied to text features as at any time step, it handles information  about  the  past  and  future  which helps improving the module accurcy.  
We feed the time domain features extracted from the audio signal into a Dense Deep Network to preserve discriminative information. Similarly, the transcribed text data are fed into BERT model.

\begin{figure}
    \centering
    \includegraphics[width=1\textwidth]{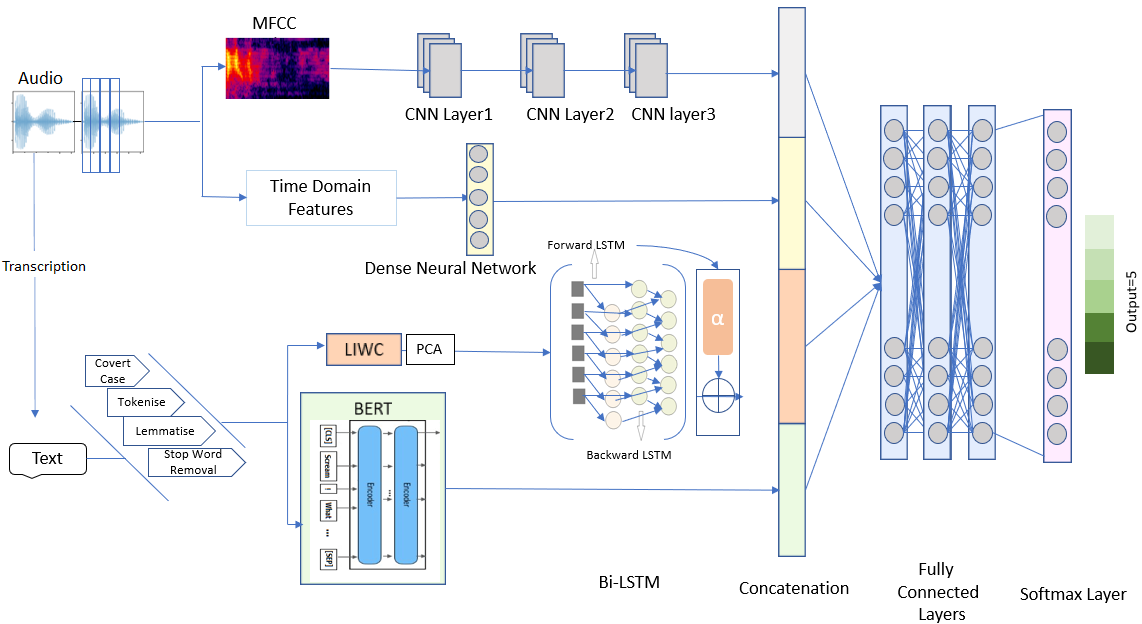}
    \caption{Framework of the proposed fusion model.  LIWC features are trained using a Bi-LSTM model with an attention layer, and MFCC Spectogram features are trained with a 3 layers 2D CNN model. Audio Time Domain features are trained using Dense Deep Network and the text is also trained using BERT. Outputs from above is then concatenated and fed into two fully connected networks (FC) followed by Softmax layer.}
    \label{fig:network}
\end{figure}

In search of the optimal model, we explored the following hyperparameter spaces: number of hidden layer (0, 1, 2, 3), dropout rate (0, 0.1, 0.5), number of hidden nodes (32, 64, 128), activation function (‘ReLU’, ‘linear’), Adam optimizer with learning rate (6.25e-3, 6.25e-4, 6.25e-5, 6.25e-6), and number of epochs (1, 5, 10). 
The best performing model comprised of 3 hidden layers, 128 hidden units in each layer (32 and 64 for CNN filters). The parameters of the three CNN layers are as follows:
\begin{equation}
    \begin{aligned}
           L_1= (K:[3,3],S:(2,2), D=0.5, F =32)\\
           L_2= (K:[3,3],S:(2,2), D=0.5, F =32)\\
           L_3= (K:[3,3],S:(2,2), D=0.5, F =64)\\  
    \end{aligned}
\end{equation}

where, $L_i$=Layer ID, K=Kernel, S=Stride , D=Dropout and F=Filter size.
A max pooling operation with a pooling size of 2 is added between the layers. Parameters settings are as follows: Dropout= 0.5 and  Activation=Rectified linear units (ReLU), FC1=128, FC2:128, FC3=128. We adopted F-Score to assess the overall performance of the models. A 10-fold cross validation strategy is adopted to evaluate the performance of the different frameworks. Each model is trained using the Adam optimizer (with learning rate of 6.25e-4) and a batch size of 32 for 1 epoch.
[-] Cross Entropy Loss function to assess loss as follows:
\begin{equation}
    L=-\sum_{i=1}^2 t_i log(p_i)
\end{equation}
where $t_i$ is the truth taking a value of 0 or 1 and the $P_i$ is the Softmax probability for the $i^th$ class.

To reduce over-fitting, a Regularization lambda of $\lambda$=0.01 is utilised  and batch normalization after every convolution layer.

Attention is defined in the following, where X is the representations of the input sentence with size (seq\textunderscore lens,embed\textunderscore dim), \(\mathbb{O}\) is the output of Bi-LSTM and is of size (seq\textunderscore slens,hidden×layers), hidden layers are the hyperparameters of BiLSTM,
$[\overrightarrow{(\mathbb{O})_f}$ and ${\overleftarrow{(\mathbb{O})_b}}$ are the forward and backward outputs of Bi-LSTM, respectively. w is the hidden length weight vector and is generated from a Fully Connected (FC) layer, W and b are the parameters of the fully connected layer, c is the weighted context and y is the final output with attention.

\begin{equation}
\begin{aligned}
    \mathbb{O},H=BiLSTM(X) \\
    \mathbb{O}=[\overrightarrow{\mathbb{O}_f},\overleftarrow{\mathbb{O}_b}]\\
    O=\overrightarrow{\mathbb{O}_f}+\overleftarrow{\mathbb{O}_b}\\
    x=WxO=b\\
    c=tanh(O)\times w\\
    y=O\times c\\   
\end{aligned}
\end{equation}\label{lstm}

\subsubsection{Multi-model Fusion}\label{sub:sub:TimeFreqDomain}
Figure \ref{fig:network} shows our overall framework for the proposed fusion model. Multimodal information integration is achieved by the  concatenation of features and embeddings from 1)BERT, 2)Bi-LSTM applied to LIWC data 3)Audio Time Domain, and 4)MFCC after applying CNN. This is then fed into three-layer Fully Connected (FC) networks, followed by a Softmax layer which assesses the type of label. The audio and text concatenation captures short-term, as well as long-term acoustic and linguistic characteristics to detect violence level in conversations.
The framework of the proposed method is shown in figure \ref{fig:network}.
To fuse the four types of information extracted from the different modalities, embeddings generated from both BERT and the BiLSTM model along with the 2D CNN representations and Audio Time Domain Dense layer are integrated. The concatenated embeddings are then passed to a three layers FC networks, which serves as a merge step. The concatenation of the embeddings is defined in the following Equation. 

\begin{equation}
\begin{aligned}
         a=Bi-LSTM(X_{embeddings})\\
         b=TimeDomainF(X_t)       \\
         c=CNN2D(X_{rep})         \\
         d=BERT(X_{embeddings})   \\
         x_{fuse}=[a,b,c,d]    
\end{aligned}
\end{equation}\label{embeddingfusion}

\subsection{Results and Discussion}\label{subsec:experimentalresults}

This section presents the experimental results achieved by the different proposed violence detection framework using multi-modal fusion.
First Table 3 shows the results of comparative Deep learning networks applied to the raw text (transcribed from the audio signal). A combination of BERT and CNN has achieved the best results with F1 score = 0.67 when applied on the raw text data comparing to BERT-LSTM alone with F1 score=0.66. This is inline with previous work on emotion recognition in which BERT yields better results when the focus is on the temporal context \cite{s20226688}.
It's not possible to compare our model performance to the previous research, because we couldn't find any similar work that aims at classifying violence based n conversations, however our model has done well comparing similar research looking at emotion recognition including depression classification. For the Audio modality alone, our model achieved an F1 score of 0.8032 comparing to depression classification from Audio and text 0.67 in \cite{8683027}.
The full compiled results of our fusion model are displayed in Table 4 when applied to single or multiple modalities for comparison. 
Audio features alone (including MFCC and Time Domain features) have achieved F1 score=0.80, while audio features with BERT (applied to text) achieved F1 score= 0.78. However, the audio features along LIWC has F1 score=0.69 which is noticeably less when applying BERT. 
When integrating all the modalities our model achieved F1 score of nearly 0.85. 

\begin{figure}
    \centering
    \includegraphics[width=0.7\textwidth]{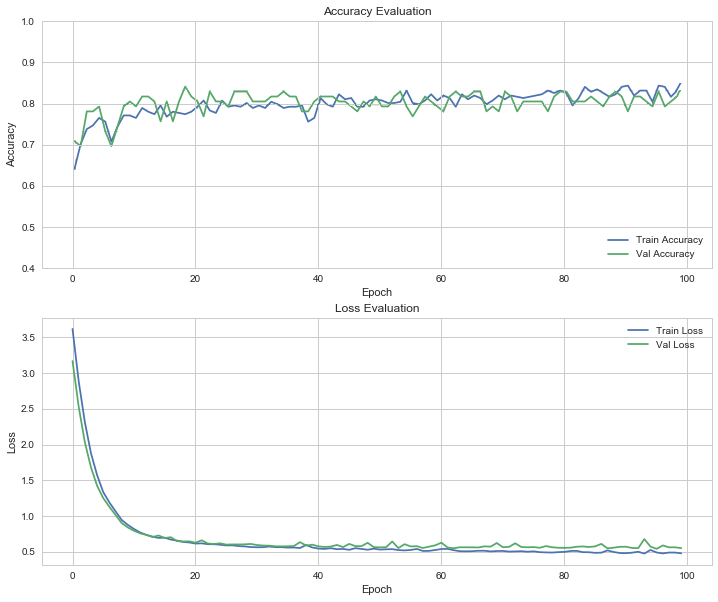}
    \caption{Train and Validation Learning Curves for Accuracy and Loss functions showing good fit.}
    \label{fig:curve}
\end{figure}

The learning curves for our model in Table 6 show good fit. See figure \ref{fig:curve} for undercurve 
\begin{table}[]
\caption{Model performance using raw text based on six combinations of CNN, LSTM, BERT and BERT (F1: F1-score).}
\centering
\begin{tabular}{p{0.25\textwidth}|p{0.20\textwidth}}
 \hline
 \textbf{Model} & \textbf{F1}\\
 \hline

 Glove-LSTM &0.5945\\
 Glove-CNN   & 0.6370\\
 Glove-NN  & 0.6602\\
 BERT-LSTM &    0.6602\\
 BERT-CNN & 0.6681\\

 \hline
\end{tabular}
\end{table}

\begin{table}[]
\caption{Overall model performance of the multimodel features, (F1: F1-score). }
\centering
\begin{tabular}{p{0.60\textwidth}|p{0.20\textwidth}}
 \hline
 \textbf{Modality} & \textbf{F1}\\
  \hline
 Baseline Model (Random Forest-all features) & 0.7469\\
  \hline
 MFCC + Time Domain features & 0.8032\\
 MFCC+Time Domain +BERT & 0.7790\\
 MFCC+Time Domain +LIWC & 0.6934\\
MFCC+Time Domain +LIWC+BERT & \textbf{0.8454}\\
 \hline
\end{tabular}
\end{table}

While this is can be considered a good outcome comparing to other machine learning and deep learning models, however the results suggest that \%15 will be miss-classified which might create some issues when a false alarm is triggered. Further work and model optimisation is required for higher accuracy levels. 
Despite the numerous existing research effort in acoustic technologies, audio based speech analysis studies and application still have a scope for further development and improvement, for examples: 

\begin{itemize}
    \item The current dataset alone requires more data to layout a better base line as most of the available abusive/violent datasets are derived from social media. Existing good examples of previous work, such as \cite{waseem-hovy:2016:N16-2} and \cite{hateoffensive} don't account for different accents as they are solemnly derived from social media. 
    \item When  examining  violence-related  data,  analysts  are not only interested in the overall violence of one particular segment or sentence but on the understanding of the type of emerging  violence-related events in a conversation. For example the word “killing” may have a violent-related orientation as in “mass  killing” while it  has a non-violent one in “killer app”. Therefore, detecting topic and violence-relatedness longitudinally should serve as a  critical  feature  in  helping  violence detectors  by  providing more  over all context.
    \item Another challenge is domain dependency as the target features are specific to the domain, it is possible for it to misinterpreting the context of the text. 
    \item Sarcasm and the tone of the person speaking is another challenge as they are subjective and could have different meanings, thus being challenging to model.   
    
\end{itemize} 

\section{Opportunities and Challenges} \label{sec:challenges}
Our future work will focus on the co-design and development of innovative edge devices (wearable or fixed) using multi-modal sensing. This means, the behavioural data required for violence detection models, need to be collected under highly variant free-living scenarios rather than  controlled settings. 
These real-world devices will incorporate embedded electronics for intervention and informed by close-to-market research. Central to this is information fusion enabled by Edge Computing (EC), embedded sensors, wireless communications, data fusion and deep learning as well as new co-design methods. 
The utilisation of the text modalities has improved the results than using the audio features by itself. However this is not only reason for the potential of text features for violence detection. In future experiments, we will look at labeling particular types of violence and text categories  which might help in providing more context to assess the level of violence. For examples, key text topics can be labelled such as hate, fight and crying which might help in determining the nature of detected violence.
In addition, we will introduce more modalities to optimise our models. For examples, movement recognition including fall detection and detection of fast or jerky movements, can help in identifying whether the violence is physical. We will consider utilising our previous research on proximity detection \cite{784104} to determine whether the victim is in close proximity to a known offender.
Since the algorithms developed in this work require high performance hardware to  such as GPU and the model  might occupy hundreds megabytes of storage space. We will implement effective strategies to accelerate and compress models including Filter Pruning \cite{Luo_2017_ICCV} and quantization base techniques such as Binary Neural Networks (BNNs) \cite{BNN2020} which enjoy a number of hardware-friendly properties including memory saving, power efficiency and significant acceleration.
Beside, since our proposed application needs to be performed in real-time, we will be looking at various techniques and tools to bring model latency down similar to \cite{VEERAMANIKANDAN2020103785}.

There might be a concern about privacy and information sharing. This work aims to share the output of the algorithms only with trusted organisations who can act immediately in case of threat or imminent danger. The users will need to give a consent to and accept to wear or use the proposed monitoring devise to protect their safety.
Making use of Edge Computing to protect privacy- by choosing a miniature edge device as the base for data processing at the point of collection the users and their supporters will have the option to keep data locally, and trigger interaction with other users without the need to share data more broadly. Future adaptations of violence detection algorithms can adopt a  multi trust layer approach to cater for the users’ needs and to allow their support group to decide how much personal information to share and with who.

\section{Conclusions} \label{sec:conclusions}
As technologies continue to evolve and victim support professionals strive for better and more advanced ways of keeping people free from harm, law enforcement bodies can incorporate new tools to better address potential threats. The use of deep multi-level feature extraction and multi-modal fusion techniques  for crime prevention can help in identifying violence or access aggression in real-time.  
The experiment showed that our fusion model can effectively mine violence information from spectral and temporal features with F1 score=0.85. The research challenges and the limitations of existing technologies suggest that the use edge computing embedded in unobtrusive wearable devices can help the society to live a better and safer life. Future work will be focused on the implementation and experimental evaluation of such wearable systems. The incorporation of additional sensing technologies will be also studied and evaluated.

\section{Acknowledgment}
The authors would like to thank the Dawes Centre for Future Crime funding for partially funding the project.



\bibliographystyle{elsarticle-num} 
\bibliography{bibliography.bib}

\begin{thebibliography}{10}
\expandafter\ifx\csname url\endcsname\relax
  \def\url#1{\texttt{#1}}\fi
\expandafter\ifx\csname urlprefix\endcsname\relax\def\urlprefix{URL }\fi
\expandafter\ifx\csname href\endcsname\relax
  \def\href#1#2{#2} \def\path#1{#1}\fi

\bibitem{Baumeister2003}
R.~F. Baumeister, B.~J. Bushman,
  \href{https://doi.org/10.1007/978-0-306-48039-3_25}{Emotions and
  Aggressiveness}, Springer Netherlands, Dordrecht, 2003, pp. 479--493.
\newblock \href {https://doi.org/10.1007/978-0-306-48039-3_25}
  {\path{doi:10.1007/978-0-306-48039-3_25}}.
\newline\urlprefix\url{https://doi.org/10.1007/978-0-306-48039-3_25}

\bibitem{allen2007patterns}
T.~Allen, S.~A. Novak, L.~L. Bench, Patterns of injuries: accident or abuse,
  Violence Against Women 13~(8) (2007) 802--816.

\bibitem{bhavan2019bagged}
A.~Bhavan, P.~Chauhan, R.~R. Shah, et~al., Bagged support vector machines for
  emotion recognition from speech, Knowledge-Based Systems 184 (2019) 104886.

\bibitem{davidson2017automated}
T.~Davidson, D.~Warmsley, M.~Macy, I.~Weber, Automated hate speech detection
  and the problem of offensive language, in: Proceedings of the International
  AAAI Conference on Web and Social Media, Vol.~11, 2017.

\bibitem{li2019improved}
Y.~Li, T.~Zhao, T.~Kawahara, Improved end-to-end speech emotion recognition
  using self attention mechanism and multitask learning., in: Interspeech,
  2019, pp. 2803--2807.

\bibitem{atmaja2019speech}
B.~T. Atmaja, M.~Akagi, Speech emotion recognition based on speech segment
  using lstm with attention model, in: 2019 IEEE International Conference on
  Signals and Systems (ICSigSys), IEEE, 2019, pp. 40--44.

\bibitem{du2019spatio}
Z.~Du, S.~Wu, D.~Huang, W.~Li, Y.~Wang, Spatio-temporal encoder-decoder fully
  convolutional network for video-based dimensional emotion recognition, IEEE
  Transactions on Affective Computing (2019).

\bibitem{hu2019video}
M.~Hu, H.~Wang, X.~Wang, J.~Yang, R.~Wang, Video facial emotion recognition
  based on local enhanced motion history image and cnn-ctslstm networks,
  Journal of Visual Communication and Image Representation 59 (2019) 176--185.

\bibitem{hajarolasvadi2020deep}
N.~Hajarolasvadi, H.~Demirel, Deep facial emotion recognition in video using
  eigenframes, IET Image Processing (2020).

\bibitem{batbaatar2019semantic}
E.~Batbaatar, M.~Li, K.~H. Ryu, Semantic-emotion neural network for emotion
  recognition from text, IEEE Access 7 (2019) 111866--111878.

\bibitem{yang2020dacnn}
C.-T. Yang, Y.-L. Chen, Dacnn: Dynamic weighted attention with multi-channel
  convolutional neural network for emotion recognition, in: 2020 21st IEEE
  International Conference on Mobile Data Management (MDM), IEEE, 2020, pp.
  316--321.

\bibitem{plaza2020improved}
F.~M. Plaza-del Arco, M.~T. Mart{\'\i}n-Valdivia, L.~A. Ure{\~n}a-L{\'o}pez,
  R.~Mitkov, Improved emotion recognition in spanish social media through
  incorporation of lexical knowledge, Future Generation Computer Systems 110
  (2020) 1000--1008.

\bibitem{joshi1991natural}
A.~K. Joshi, Natural language processing, Science 253~(5025) (1991) 1242--1249.

\bibitem{mossie2020vulnerable}
Z.~Mossie, J.-H. Wang, Vulnerable community identification using hate speech
  detection on social media, Information Processing \& Management 57~(3) (2020)
  102087.

\bibitem{sharma2018nlp}
H.~K. Sharma, K.~Kshitiz, et~al., Nlp and machine learning techniques for
  detecting insulting comments on social networking platforms, in: 2018
  International Conference on Advances in Computing and Communication
  Engineering (ICACCE), IEEE, 2018, pp. 265--272.

\bibitem{Giannakopoulo06}
T.~Giannakopoulos, D.~Kosmopoulos, A.~Aristidou, S.~Theodoridis, Violence
  content classification using audio features, in: G.~Antoniou, G.~Potamias,
  C.~Spyropoulos, D.~Plexousakis (Eds.), Advances in Artificial Intelligence,
  Springer Berlin Heidelberg, Berlin, Heidelberg, 2006, pp. 502--507.

\bibitem{pennebaker2001linguistic}
J.~W. Pennebaker, M.~E. Francis, R.~J. Booth, Linguistic inquiry and word
  count: Liwc 2001, Mahway: Lawrence Erlbaum Associates 71~(2001) (2001) 2001.

\bibitem{devlin2018bert}
J.~Devlin, M.-W. Chang, K.~Lee, K.~Toutanova, Bert: Pre-training of deep
  bidirectional transformers for language understanding, arXiv preprint
  arXiv:1810.04805 (2018).

\bibitem{medhat2014sentiment}
W.~Medhat, A.~Hassan, H.~Korashy, Sentiment analysis algorithms and
  applications: A survey, Ain Shams engineering journal 5~(4) (2014)
  1093--1113.

\bibitem{watanabe2018hate}
H.~Watanabe, M.~Bouazizi, T.~Ohtsuki, Hate speech on twitter: A pragmatic
  approach to collect hateful and offensive expressions and perform hate speech
  detection, IEEE access 6 (2018) 13825--13835.

\bibitem{cano2013weakly}
A.~E. Cano~Basave, Y.~He, K.~Liu, J.~Zhao, A weakly supervised bayesian model
  for violence detection in social media (2013).

\bibitem{zhang2018deep}
L.~Zhang, S.~Wang, B.~Liu, Deep learning for sentiment analysis: A survey,
  Wiley Interdisciplinary Reviews: Data Mining and Knowledge Discovery 8~(4)
  (2018) e1253.

\bibitem{poria2017review}
S.~Poria, E.~Cambria, R.~Bajpai, A.~Hussain, A review of affective computing:
  From unimodal analysis to multimodal fusion, Information Fusion 37 (2017)
  98--125.

\bibitem{gemmeke2017audio}
J.~F. Gemmeke, D.~P. Ellis, D.~Freedman, A.~Jansen, W.~Lawrence, R.~C. Moore,
  M.~Plakal, M.~Ritter, Audio set: An ontology and human-labeled dataset for
  audio events, in: 2017 IEEE international conference on acoustics, speech and
  signal processing (ICASSP), IEEE, 2017, pp. 776--780.

\bibitem{giannakopoulos2006violence}
T.~Giannakopoulos, D.~Kosmopoulos, A.~Aristidou, S.~Theodoridis, Violence
  content classification using audio features, in: Hellenic Conference on
  Artificial Intelligence, Springer, 2006, pp. 502--507.

\bibitem{ortigosa2014sentiment}
A.~Ortigosa, J.~M. Mart{\'\i}n, R.~M. Carro, Sentiment analysis in facebook and
  its application to e-learning, Computers in human behavior 31 (2014)
  527--541.

\bibitem{naf2019sentiment}
M.~Z. Naf'an, A.~A. Bimantara, A.~Larasati, E.~M. Risondang, N.~A.~S. Nugraha,
  Sentiment analysis of cyberbullying on instagram user comments, Journal of
  Data Science and Its Applications 2~(1) (2019) 38--48.

\bibitem{severyn2015twitter}
A.~Severyn, A.~Moschitti, Twitter sentiment analysis with deep convolutional
  neural networks, in: Proceedings of the 38th International ACM SIGIR
  Conference on Research and Development in Information Retrieval, 2015, pp.
  959--962.

\bibitem{HUSSEIN2018330}
D.~M. E.-D.~M. Hussein,
  \href{https://www.sciencedirect.com/science/article/pii/S1018363916300071}{A
  survey on sentiment analysis challenges}, Journal of King Saud University -
  Engineering Sciences 30~(4) (2018) 330--338.
\newblock \href {https://doi.org/https://doi.org/10.1016/j.jksues.2016.04.002}
  {\path{doi:https://doi.org/10.1016/j.jksues.2016.04.002}}.
\newline\urlprefix\url{https://www.sciencedirect.com/science/article/pii/S1018363916300071}

\bibitem{CHATURVEDI201865}
I.~Chaturvedi, E.~Cambria, R.~E. Welsch, F.~Herrera,
  \href{https://www.sciencedirect.com/science/article/pii/S1566253517303901}{Distinguishing
  between facts and opinions for sentiment analysis: Survey and challenges},
  Information Fusion 44 (2018) 65--77.
\newblock \href {https://doi.org/https://doi.org/10.1016/j.inffus.2017.12.006}
  {\path{doi:https://doi.org/10.1016/j.inffus.2017.12.006}}.
\newline\urlprefix\url{https://www.sciencedirect.com/science/article/pii/S1566253517303901}

\bibitem{zhao2016automatic}
R.~Zhao, A.~Zhou, K.~Mao, Automatic detection of cyberbullying on social
  networks based on bullying features, in: Proceedings of the 17th
  international conference on distributed computing and networking, 2016, pp.
  1--6.

\bibitem{chen2011detecting}
Y.~Chen, Detecting offensive language in social medias for protection of
  adolescent online safety (2011).

\bibitem{chen2019complementary}
F.~Chen, Z.~Luo, Y.~Xu, D.~Ke, Complementary fusion of multi-features and
  multi-modalities in sentiment analysis, arXiv preprint arXiv:1904.08138
  (2019).

\bibitem{chen2012detecting}
Y.~Chen, Y.~Zhou, S.~Zhu, H.~Xu, Detecting offensive language in social media
  to protect adolescent online safety, in: 2012 International Conference on
  Privacy, Security, Risk and Trust and 2012 International Confernece on Social
  Computing, IEEE, 2012, pp. 71--80.

\bibitem{cambria2017sentiment}
E.~Cambria, S.~Poria, A.~Gelbukh, M.~Thelwall, Sentiment analysis is a big
  suitcase, IEEE Intelligent Systems 32~(6) (2017) 74--80.

\bibitem{warner2012detecting}
W.~Warner, J.~Hirschberg, Detecting hate speech on the world wide web, in:
  Proceedings of the second workshop on language in social media, 2012, pp.
  19--26.

\bibitem{burnap2015cyber}
P.~Burnap, M.~L. Williams, Cyber hate speech on twitter: An application of
  machine classification and statistical modeling for policy and decision
  making, Policy \& Internet 7~(2) (2015) 223--242.

\bibitem{peng2019transfer}
Y.~Peng, S.~Yan, Z.~Lu, Transfer learning in biomedical natural language
  processing: an evaluation of bert and elmo on ten benchmarking datasets,
  arXiv preprint arXiv:1906.05474 (2019).

\bibitem{howard2018universal}
J.~Howard, S.~Ruder, Universal language model fine-tuning for text
  classification, arXiv preprint arXiv:1801.06146 (2018).

\bibitem{qu2020text}
Y.~Qu, P.~Liu, W.~Song, L.~Liu, M.~Cheng, A text generation and prediction
  system: Pre-training on new corpora using bert and gpt-2, in: 2020 IEEE 10th
  International Conference on Electronics Information and Emergency
  Communication (ICEIEC), IEEE, 2020, pp. 323--326.

\bibitem{alatawi2021detecting}
H.~S. Alatawi, A.~M. Alhothali, K.~M. Moria, Detecting white supremacist hate
  speech using domain specific word embedding with deep learning and bert, IEEE
  Access 9 (2021) 106363--106374.

\bibitem{kaushik2013sentiment}
L.~Kaushik, A.~Sangwan, J.~H. Hansen, Sentiment extraction from natural audio
  streams, in: 2013 IEEE International Conference on Acoustics, Speech and
  Signal Processing, IEEE, 2013, pp. 8485--8489.

\bibitem{maghilnan2017sentiment}
S.~Maghilnan, M.~R. Kumar, Sentiment analysis on speaker specific speech data,
  in: 2017 International Conference on Intelligent Computing and Control
  (I2C2), IEEE, 2017, pp. 1--5.

\bibitem{luo2019audio}
Z.~Luo, H.~Xu, F.~Chen, Audio sentiment analysis by heterogeneous signal
  features learned from utterance-based parallel neural network, in: AffCon@
  AAAI, 2019.

\bibitem{li2019acoustic}
B.~Li, D.~Dimitriadis, A.~Stolcke, Acoustic and lexical sentiment analysis for
  customer service calls, in: ICASSP 2019-2019 IEEE International Conference on
  Acoustics, Speech and Signal Processing (ICASSP), IEEE, 2019, pp. 5876--5880.

\bibitem{rane2020audio}
A.~L. Rane, A.~R. Kshatriya, Audio opinion mining and sentiment analysis of
  customer product or services reviews, in: ICDSMLA 2019, Springer, 2020, pp.
  282--293.

\bibitem{yenigalla2018speech}
P.~Yenigalla, A.~Kumar, S.~Tripathi, C.~Singh, S.~Kar, J.~Vepa, Speech emotion
  recognition using spectrogram \& phoneme embedding., in: Interspeech, 2018,
  pp. 3688--3692.

\bibitem{mikolov2013efficient}
T.~Mikolov, K.~Chen, G.~Corrado, J.~Dean, Efficient estimation of word
  representations in vector space, arXiv preprint arXiv:1301.3781 (2013).

\bibitem{povolny2016multimodal}
F.~Povolny, P.~Matejka, M.~Hradis, A.~Popkov{\'a}, L.~Otrusina, P.~Smrz,
  I.~Wood, C.~Robin, L.~Lamel, Multimodal emotion recognition for avec 2016
  challenge, in: Proceedings of the 6th International Workshop on Audio/Visual
  Emotion Challenge, 2016, pp. 75--82.

\bibitem{perez2013utterance}
V.~P{\'e}rez-Rosas, R.~Mihalcea, L.-P. Morency, Utterance-level multimodal
  sentiment analysis, in: Proceedings of the 51st Annual Meeting of the
  Association for Computational Linguistics (Volume 1: Long Papers), 2013, pp.
  973--982.

\bibitem{wollmer2013youtube}
M.~W{\"o}llmer, F.~Weninger, T.~Knaup, B.~Schuller, C.~Sun, K.~Sagae, L.-P.
  Morency, Youtube movie reviews: Sentiment analysis in an audio-visual
  context, IEEE Intelligent Systems 28~(3) (2013) 46--53.

\bibitem{poria2015deep}
S.~Poria, E.~Cambria, A.~Gelbukh, Deep convolutional neural network textual
  features and multiple kernel learning for utterance-level multimodal
  sentiment analysis, in: Proceedings of the 2015 conference on empirical
  methods in natural language processing, 2015, pp. 2539--2544.

\bibitem{yoon2018multimodal}
S.~Yoon, S.~Byun, K.~Jung, Multimodal speech emotion recognition using audio
  and text, in: 2018 IEEE Spoken Language Technology Workshop (SLT), IEEE,
  2018, pp. 112--118.

\bibitem{buitelaar2018mixedemotions}
P.~Buitelaar, I.~D. Wood, S.~Negi, M.~Arcan, J.~P. McCrae, A.~Abele, C.~Robin,
  V.~Andryushechkin, H.~Ziad, H.~Sagha, et~al., Mixedemotions: An open-source
  toolbox for multimodal emotion analysis, IEEE Transactions on Multimedia
  20~(9) (2018) 2454--2465.

\bibitem{sahu2019multimodal}
G.~Sahu, Multimodal speech emotion recognition and ambiguity resolution, arXiv
  preprint arXiv:1904.06022 (2019).

\bibitem{pereira2016fusing}
M.~H.~R. Pereira, F.~L.~C. P{\'a}dua, A.~C.~M. Pereira, F.~Benevenuto, D.~H.
  Dalip, Fusing audio, textual, and visual features for sentiment analysis of
  news videos, in: Tenth International AAAI Conference on Web and Social Media,
  2016.

\bibitem{youtube}
G.~Atkins, \href{https://www.youtube.com/watch?v=6RbdaqWa8qY}{Eastenders.
  gray’s torture \& abuse of chantelle atkins full story} (2020).
\newline\urlprefix\url{https://www.youtube.com/watch?v=6RbdaqWa8qY}

\bibitem{sood2012using}
S.~O. Sood, J.~Antin, E.~Churchill, Using crowdsourcing to improve profanity
  detection, in: 2012 AAAI Spring Symposium Series, 2012.

\bibitem{macavaney2019hate}
S.~MacAvaney, H.-R. Yao, E.~Yang, K.~Russell, N.~Goharian, O.~Frieder, Hate
  speech detection: Challenges and solutions, PloS one 14~(8) (2019) e0221152.

\bibitem{waseem2016hateful}
Z.~Waseem, D.~Hovy, Hateful symbols or hateful people? predictive features for
  hate speech detection on twitter, in: Proceedings of the NAACL student
  research workshop, 2016, pp. 88--93.

\bibitem{mozafari2020hate}
M.~Mozafari, R.~Farahbakhsh, N.~Crespi, Hate speech detection and racial bias
  mitigation in social media based on bert model, PloS one 15~(8) (2020)
  e0237861.

\bibitem{Pennebaker15}
R.~James~W., Boyd, Jordan, The development and psychometric properties of
  liwc2015 (2015).

\bibitem{Jolliffe90}
I.~T. Jolliffe, Principal component analysis: A beginner's guide i:
  Introduction and application, Weather (1990).
\newblock \href
  {https://doi.org/https://doi.org/10.1002/j.1477-8696.1990.tb05558.x}
  {\path{doi:https://doi.org/10.1002/j.1477-8696.1990.tb05558.x}}.

\bibitem{784104}
R.~Vergin, D.~O'Shaughnessy, A.~Farhat, Generalized mel frequency cepstral
  coefficients for large-vocabulary speaker-independent continuous-speech
  recognition, IEEE Transactions on Speech and Audio Processing 7~(5) (1999)
  525--532.
\newblock \href {https://doi.org/10.1109/89.784104}
  {\path{doi:10.1109/89.784104}}.

\bibitem{9231008}
F.~M. Shah, F.~Ahmed, S.~K. Saha~Joy, S.~Ahmed, S.~Sadek, R.~Shil, M.~H. Kabir,
  Early depression detection from social network using deep learning
  techniques, in: 2020 IEEE Region 10 Symposium (TENSYMP), 2020, pp. 823--826.
\newblock \href {https://doi.org/10.1109/TENSYMP50017.2020.9231008}
  {\path{doi:10.1109/TENSYMP50017.2020.9231008}}.

\bibitem{s20226688}
S.~Lee, D.~K. Han, H.~Ko,
  \href{https://www.mdpi.com/1424-8220/20/22/6688}{Fusion-convbert: Parallel
  convolution and bert fusion for speech emotion recognition}, Sensors 20~(22)
  (2020).
\newblock \href {https://doi.org/10.3390/s20226688}
  {\path{doi:10.3390/s20226688}}.
\newline\urlprefix\url{https://www.mdpi.com/1424-8220/20/22/6688}

\bibitem{8683027}
G.~Lam, H.~Dongyan, W.~Lin, Context-aware deep learning for multi-modal
  depression detection, in: ICASSP 2019 - 2019 IEEE International Conference on
  Acoustics, Speech and Signal Processing (ICASSP), 2019, pp. 3946--3950.
\newblock \href {https://doi.org/10.1109/ICASSP.2019.8683027}
  {\path{doi:10.1109/ICASSP.2019.8683027}}.

\bibitem{waseem-hovy:2016:N16-2}
Z.~Waseem, D.~Hovy, \href{http://www.aclweb.org/anthology/N16-2013}{Hateful
  symbols or hateful people? predictive features for hate speech detection on
  twitter}, in: Proceedings of the NAACL Student Research Workshop, Association
  for Computational Linguistics, San Diego, California, 2016, pp. 88--93.
\newline\urlprefix\url{http://www.aclweb.org/anthology/N16-2013}

\bibitem{hateoffensive}
T.~Davidson, D.~Warmsley, M.~Macy, I.~Weber, Automated hate speech detection
  and the problem of offensive language, in: Proceedings of the 11th
  International AAAI Conference on Web and Social Media, ICWSM '17, 2017, pp.
  512--515.

\bibitem{Luo_2017_ICCV}
J.-H. Luo, J.~Wu, W.~Lin, Thinet: A filter level pruning method for deep neural
  network compression, in: Proceedings of the IEEE International Conference on
  Computer Vision (ICCV), 2017.

\bibitem{BNN2020}
H.~Qin, R.~Gong, X.~Liu, X.~Bai, J.~Song, N.~Sebe,
  \href{http://dx.doi.org/10.1016/j.patcog.2020.107281}{Binary neural networks:
  A survey}, Pattern Recognition 105 (2020) 107281.
\newblock \href {https://doi.org/10.1016/j.patcog.2020.107281}
  {\path{doi:10.1016/j.patcog.2020.107281}}.
\newline\urlprefix\url{http://dx.doi.org/10.1016/j.patcog.2020.107281}

\bibitem{VEERAMANIKANDAN2020103785}
Veeramanikandan, S.~Sankaranarayanan, J.~J. Rodrigues, V.~Sugumaran, S.~Kozlov,
  \href{https://www.sciencedirect.com/science/article/pii/S0952197620301780}{Data
  flow and distributed deep neural network based low latency iot-edge
  computation model for big data environment}, Engineering Applications of
  Artificial Intelligence 94 (2020) 103785.
\newblock \href
  {https://doi.org/https://doi.org/10.1016/j.engappai.2020.103785}
  {\path{doi:https://doi.org/10.1016/j.engappai.2020.103785}}.
\newline\urlprefix\url{https://www.sciencedirect.com/science/article/pii/S0952197620301780}

\end{thebibliography}





\end{document}